\documentclass{sig-alternate}
\usepackage{graphicx}
\begin{document}
%

\title{Applying Semantic Web Technologies for Improving the Visibility of Tourism Data}
%
%
%
%
%
\numberofauthors{1} 
%
\author{
%
%
Fayrouz Soualah-Alila, Cyril Faucher, Fr\'ed\'eric Bertrand, \\Micka\"el Coustaty, Antoine Doucet\\
       \affaddr{L3I laboratory}\\
       \affaddr{La Rochelle University}\\
       \affaddr{Michel Cr\'epeau avenue, 17000 La Rochelle, France}\\
       \email{\normalsize{\{fayrouz.soualah-alila, cyril.faucher, fbertran, mickael.coustaty, antoine.doucet\}@univ-lr.fr}}
}


\maketitle
\begin{abstract}
Tourism industry is an extremely information-intensive, complex and dynamic activity.
It can benefit from semantic Web technologies, due to the significant heterogeneity of information sources and the high volume of on-line data.
The management of semantically diverse annotated tourism data is facilitated by ontologies that provide methods and standards, which allow flexibility and more intelligent access to on-line data.
This paper provides a description of some of the early results of the Tourinflux project which aims to apply semantic Web technologies to support tourist actors in effectively finding and publishing information on the Web.

\end{abstract}




\section{Introduction}

\subsection{Context}
The tourism is viewed as information intensive industry and a highly dynamic and changing domain where information plays an important role for decision and action making \cite{Inkpen1998}.
Tourism represents a significant part of the global economy, more than 9\% of the global GDP (US \$7 trillion) \cite{WTTC2014}.
It becomes increasingly difficult for policy makers, on the one hand to guide choices in order to boost and make their territory attractive, and on the other hand to analyze the comments and the vision of the territory by users.
As the World Wide Web has changed people's daily life, it has significantly influenced the way of information gathering and exchanging in the tourism area, with the intensive use of social networks and Web sites specialized in e-tourism (TripAdvisor, Booking.com, etc.).
Along with the Web technologies, tourism actors such as tour operators need knowledge about travel, hotels, markets, etc.
Indeed, information like hotel prices, hotel localization, offers, market tendencies are used to improve their service quality by enhancing employee's knowledge about customer's preferences.
Other actors such as experts from tourism industry are also making use of the Web to promote their territory, and so to attract more visitors.\\
However, the gaps of the existing Web technologies arise two main topics:\\
First, tourism domain is characterized by a significant heterogeneity of information sources and by a high volume of on-line data.
Data related to tourism is produced by different experts (travel agents, tourist offices, etc.) and by visitors, constituting semantically heterogeneous data, often incomplete and inconsistent.
This data could be composed of information related to tourism objects (hotels, concert, restaurant, etc.), temporal information or opinions.
There already exist different taxonomies and catalogues which are designed and used internally by tourism actors to help them to manage heterogeneous tourism data.
Efforts are now made to generate standards to facilitate inter and intra tourism data exchange.\\
Second, the information contained on Web pages are originally designed to be human-readable, and so, most of information currently available on the Web are kept in large collections of textual documents.
As the Web grows in size and complexity, there is an increasing need for automating some of the time consuming tasks such as information searching, extraction and interpretation.
Semantic Web technologies are an answer to this issue by proposing some novel and sophisticated question answering systems. In this way, ontologies provides a formal framework to organize data and to browse, search or access this information \cite{Bruijn2008}.

\subsection{Tourinflux}
This work takes place under the Tourinflux\footnote{http://tourinflux.univ-lr.fr/} project which aims at providing the tourism industry with a set of tools (1) allowing them to handle both their internal data, and the information available on the Web, and (2) allowing to improve the display information available about their territory on the Web.
This should allow them to improve their decision process and subsequently, their effectiveness.\\
This paper presents some preliminary results of Tourinflux project.
The paper aim to describe the development of an ontology based model for tourism domain.
It is aimed primarily to present some aspects of knowledge management:
Section 2 reviews the related ontologies developed in the previous projects.
In particular, we describe how an ontology can be used to improve the quality of users Web search and data publication.
Section 2 also provides description of current tourism standards and a brief background on a particular ontology: the Schema.org\footnote{http://schema.org/} model.
In an early stage of our project, a partial ontology called TIFSem (Semantic TourInFrance) for the tourism was created, using Prot\'{e}g\'{e} \footnote{http://protege.stanford.edu/} and the Web Ontology Language \footnote{http://www.w3.org/2001/sw/wiki/OWL} (OWL).
A partial view of TIFSem ontology developed will be presented in section 3 with some example of queries.
We notice that this work is in progress; we are still gathering new concepts for its taxonomy and new axioms in order to validate them with experts of the domain.
Finally, conclusions and outlook are provided in section 4.

\section{State of art}

\subsection{Tourism ontologies}
In the domain of tourism, several researches have focused on the design of ontologies.
Several available tourism ontologies show the current status of the efforts:\\
The OTA (OpenTravelAlliance) specification defines XML Schema documents corresponding to events and activities in various travel sectors \cite{OTA2000}.\\
The Harmonise ontology is specialised to address interoperability problems focusing on data exchange \cite{Dell2002}.
Harmonise is based on mapping different tourism ontologies by using a mediating ontology.
This central Harmonise ontology is represented in RDF and focuses mainly on tourism events and accommodation types.\\
The Hi-Touch ontology models tourism destinations and their associated documentations \cite{Legrand2004}.
The ontology classifies tourist objects, which are linked together in a network by semantic relationships. The top-level classes of the Hi-Touch ontology are documents (any kind of documentation about a
tourism product), objects (the tourism objects themselves) and publication (a document created from the results of a query).\\
The QALL-ME ontology provides a model to describe tourism destinations, sites, events and transportation \cite{Ou2008} and aims at establishing a shared infrastructure for multilingual and multimodal question answering.
The Tourpedia catalogue describes places (points, restaurants, accommodations and interests) related to different locations (Amsterdam, London, Paris, Rome, etc.) \cite{Cresci2014}.\\
These models focus on different areas of tourism domain but there does not exist one single ontology which matches all the needs of different tourism related applications.
We then propose in section 3 an ontology to globally describe tourism information mixing heterogeneous content.
Concepts included in the defined ontology will allow us to annotate information sources on tourism.
Then the enriched information can be used: (1) from the user side, to match tailored package holidays to client preferences and (2) from tourism experts' side, to analyze and better manage on-line data about their territory.

\subsection{Tourism standards}
The interoperability of Tourism Information Systems (TIS) is a major challenge for the development of tourism domain.
Several national, European and international institutional initiatives have proposed different standards to meet the specific needs of tourism professionals.
In France, professionals have adopted the TourInFrance standard (TIF).
Since its creation in 1999 by the tourism ministry, TourInFrance is used today by more than 3000 tourist offices in France, by Departmental Tourist Committees (DTC) and by different tour operators, to facilitate data exchange between these different actors.
In 2004, the TourInFrance Technical Group (TIFTG) approved the new version of the standard, TIF V3.
In this version, the standard has evolved towards XML technologies to facilitate publishing information on the Web and information exchange between systems, and is accompanied by several thesaurus.
The last version of TIF can be downloaded from the following Web site: http://tourinfrance.free.fr/.
Since 2005, this standard stopped evolving.
As a result, tourism professionals have adapted the standard to their own needs and proposed their own evolution in an unorganized way, producing TIS that are not really interoperable among themselves and that cannot directly be shared using international standard.
Accordingly, the exploitation of tourism information is trapped in its own territory, and so it is impossible to aggregate these information.\\
To illustrate this evolution of TIF we show in the following example how the standard was transformed.
The figures 1 and 2 show different syntaxes of TIF to describe an address; data represented in the figure 1 respect the standard TIF V3, while data represented in the figure 2 is structured under a derived version of TIF (different tags syntaxes, new tags added, etc.).\\
To overcome these limitations we want to evolve the TIF standard to share the knowledge it represents and to ensure data interoperability, by applying the concept of ontology to represent the standard terminology \cite{Bittner2005}.

\begin{figure}[h!]
\centering
\includegraphics[scale=0.35]{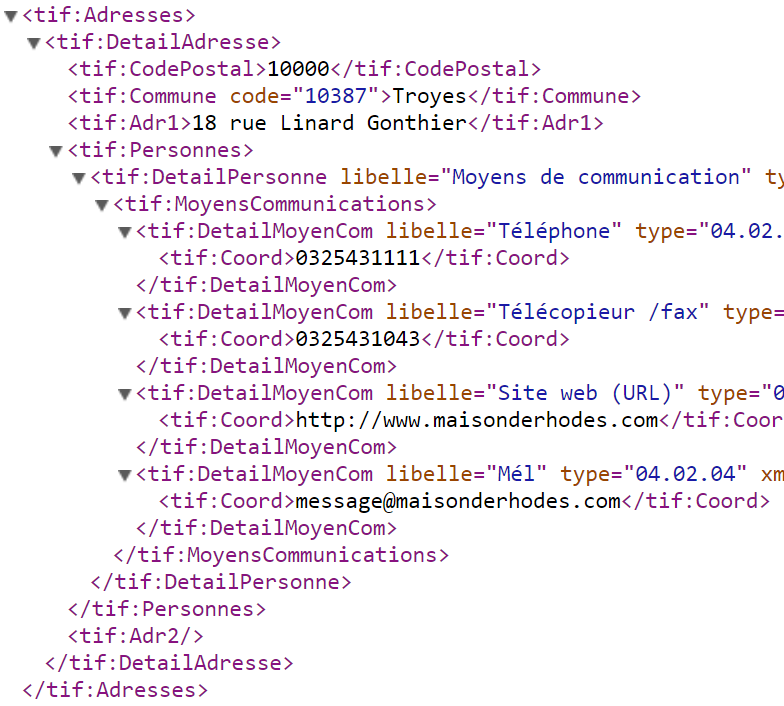}
\caption{\scriptsize{Data respecting the standard TIF V3}}
\end{figure}

\begin{figure}[h!]
\centering
\includegraphics[scale=0.38]{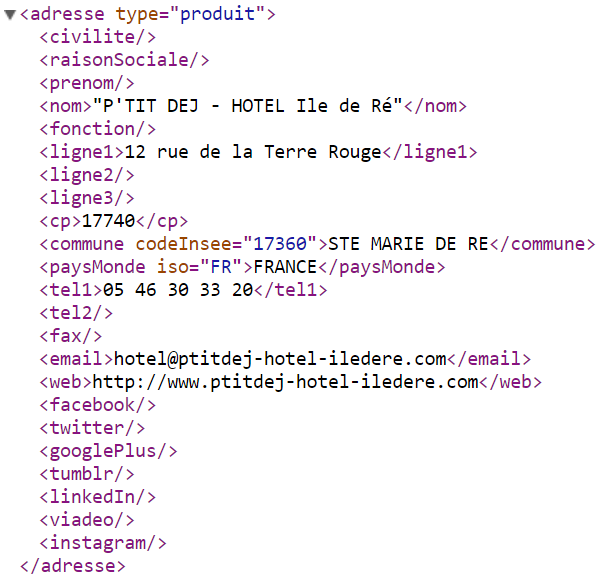}
\caption{\scriptsize{Data structured under a derived version of TIF}}
\end{figure}

\subsection{Semantic Web standards}
TIF standard is unable to easily share and interoperate with global Web standards.
Since few years, an initiative led by Bing, Google and Yahoo tends to become a de-facto norm to easily share semantic content.
Launched in 2011, Schema.org aims to create and to support a common set of schemas for structured data markup Web pages.
The purpose of Schema.org is to provide a collection of schemas (HTML tags) that webmasters can use to markup HTML pages in ways recognized by major search providers, and that can also be used for structured data interoperability (RDFa, JSON-LD, etc.).
Upon using these tags in a website, search engines can understand the intention of using resource (text, image, video) in that website better; which in turn would return better search results when a user is looking for a specific resource through a search engine \cite{Toma2014}.
Schema.org is a very large vocabulary counting hundreds of terms from multiple domains (the full specification of Schema.org is available at https://schema.org/docs/full.html).\\
In our approach we want to spread enriched semantic tourist data that can be easily indexed by search engines.
One solution would be to use directly the Schema.org ontology which provides good indexing by search engines and which is easy to implement.
The disadvantages of this solution is that we lose precision in the tourist data of the TIF model, in addition, using directly Schema.org can cause economic problems because a large number of tourist offices already use TIF or derived versions.
A second solution would be to realize an ontology by matching terms of TIF with terms of Schema.org by using OWL relations, and work with Schema.org community to extend the schema, either formally by adding new terms or informally by defining how Schema.org can be combined with some additional vocabulary terms.
We can thus produce more accurate information than the terms of Schema.org, in addition with a TIF/Schema.org correspondence, producing an interoperable model becomes possible from XML TIF sources.\\
This method is more complex to implement because it requires to connect each term of TIF with a terms of Schema.org.
The difficulty is not the transformation of XML TIF to OWL, but searching correspondence between the terms of TIF and those of Schema.org, and seeking what is lacking in each model.
We suggest to consider the second solution, which presents a better compromise.

\begin{figure}[!h]
\centering
\includegraphics[scale=0.33]{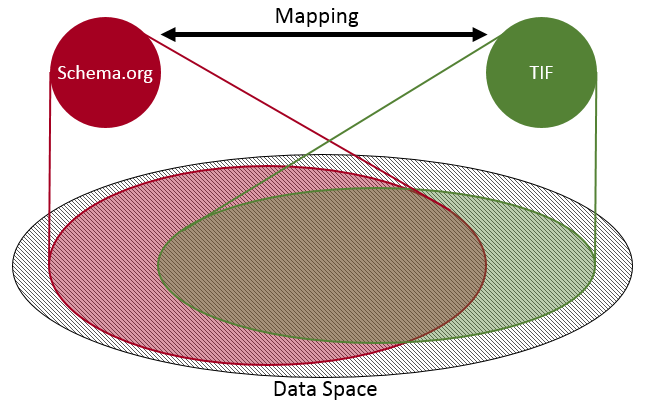}
\caption{\scriptsize{Schema.org/TIF mapping}}
\end{figure}

We consider the Schema.org as the extension core of the TIF standard to improve the online visibility of touristic service providers.
Schema.org includes many of the TIF terms.
Of course not all of the Schema.org terms are relevant for the tourism domain.
The relevant terms are those that belong to the categories Hotels, Food Establishments, Events, Trips, Place of Interest and News.
These tourism terms are distributed in different trees of the schema.
For example:
\vspace{-2mm}
\begin{itemize}
\item Thing $\rightarrow$ Place $\rightarrow$ LocalBusiness $\rightarrow$ LodgingBusiness:
\begin{itemize}
\vspace{-2mm}
\item Hostel, Hotel, Motel, etc.
\vspace{-2mm}
\end{itemize}

%
\item Thing $\rightarrow$ Event:
\begin{itemize}
\vspace{-2mm}
\item MusicEvent, SocialEvent, SportsEvent, etc.
\vspace{-2mm}
\end{itemize}
\end{itemize}

Schema.org offers other interesting terms such as customer reviews and reservation:
\begin{itemize}
\vspace{-2mm}
\item Thing $\rightarrow$ Action $\rightarrow$ AssessAction $\rightarrow$ ReviewAction
\vspace{-2mm}
\item Thing $\rightarrow$ Intangible $\rightarrow$ Reservation:
\begin{itemize}
\vspace{-2mm}
\item EventReservation, FoodEstablishmentReservation, LodgingReservation, etc.
\vspace{-2mm}
\end{itemize}
\end{itemize}

In addition, several Linked data initiatives have enabled the development of mappings between Schema.org and unavoidable semantic resources such as DBPedia, Dublin Core, FOAF, GoodRelations, SIOC and WordNet (full mappings of Web data vocabularies to Schema.org terms are available at http://schema.rdfs.org/mappings.html).
We are interested in establishing the mappings at two levels.
One will be a mapping between concepts which have the same name or approximate name, and between concepts which have the same role.
Then taking in to account the mapped concepts (equivalentClass, subClassOf), we need to match one property from the Schema.org term with another property from a concept in TIF (equivalentProperty, subPropertyOf).

\section{Proposed approach}
The tourist information can be used in different areas and different contexts.
It is important that this information have to be modular, so that the elements describing the same entity can be used separately, hence the notion of a modular information represented by the IO.
An \textit{InformationObject} (IO) is a modular entity, that can be used and re-used to support tourism activities.
An IO can be a hotel, a restaurant, an event, etc., or any other resource that can be attached to a tourism activity.
The idea of IO is to create an entity that is:

\begin{itemize}
\vspace{-2mm}
\item interoperable: can plug-and-play with any TIS,
\vspace{-2mm}
\item reusable: can be used or adapted for different use cases,
\vspace{-5mm}
\item accessible: can be stored a way that allows for easy searchability,
\vspace{-2mm}
\item manageable: can be tracked and updated over time.
\end{itemize}

\renewcommand{\arraystretch}{1.3}
\begin{table*}
\centering
\caption{Core concepts of the TIFSem ontology}
\begin{tabular}{|l|p{12.4cm}|}
\hline
\textbf{CORE CONCEPTS} & \textbf{Description}\\
\hline
DUBLIN CORE & The Dublin Core is a set of vocabulary terms that can be used to describe Web resources (video, images, Web pages, etc.).\\
\hline
UPDATE & Information about the updates of a resource, and information used to represent aspects of the evolution of the resource.\\
\hline
MULTIMEDIA & Information about multimedia materials associated to the described resource.\\
\hline
CONTACTS & Information about ways to be in relation with a resource, or a person responsible for this resource.\\
\hline
LEGAL INFORMATION & Information to formally identify an entity and its activities.\\
\hline
CLASSIFICATIONS & Information used to qualify the resource described and identify the quality of services associated.\\
\hline
RELATED SERVICES & Information to reference other resources.\\
\hline
GEOLOCATIONS & Information to locate and describe the environment a resource.\\
\hline
PERIODS & Information about an opening period, a closing period, a reservation period, etc.\\
\hline
CUSTOMERS & Customer information.\\
\hline
LANGUAGES & Languages spoken in the described resource.\\
\hline
RESERVATION MODES & Information about the reservation for the described resource, namely, who to contact customers and if the reservation is required.\\
\hline
PRICES & Services' price of the resource described, as well as the different means of payment.\\
\hline
CAPACITY & The capacity of the resource.\\
\hline
OFFERS SERVICES & Information about the services available within the described resource or nearby.\\
\hline
ADDITIONAL DESCRIPTION & Additional information about the described resource.\\
\hline
ITINERARIES & Activities associated to the described resource such as hiking.\\
\hline
SCHEDULES & Information about the availability status of services, over defined periods.\\
\hline
\end{tabular}
\end{table*}

In TIF, a typical IO is composed by 18 parts called granules (or semantic units). These granules form the main core of the TIFSem ontology.
We introduce then the TIFSem ontology for the tourist domain, which contains concepts and relations for tourist resources. It consists of 19 main concepts of tourist information described by the table 1.
The main concepts of that ontology have sub-concepts pertinent to the local area, and for all sub-concepts they have \textit{is\_a} relationships.
Those sub-concepts have properties, restrictions and semantic relations among other concepts.

The division of these main concepts into sub-concepts, was also guided by different data sources.
Indeed, in order to elaborate the TIFSem ontology, we have consulted different kinds of sources to enable the understanding and the collecting of concepts related to the specialized domain of tourism, and the corresponding vocabulary.
Sources coming from Departmental Tourism Committee of the Charente-Maritime\footnote{http://www.charente-maritime.org/} (CDT17) and Departmental Tourism Committee of the Aube\footnote{http://www.aube-champagne.com/} (CDT10) were consulted.

\begin{figure*}[!h]
\centering
\includegraphics[scale=0.45]{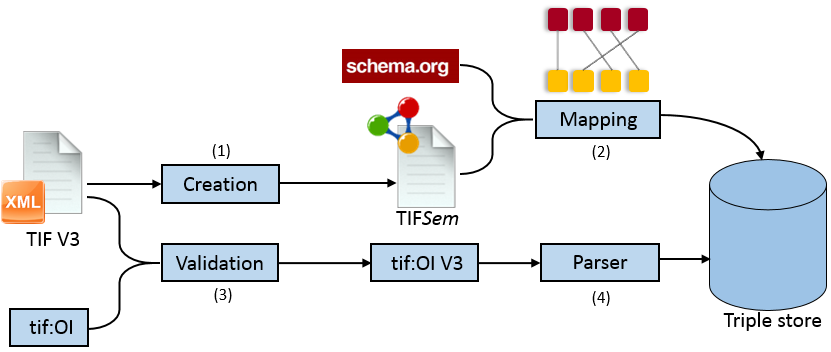}
\caption{\scriptsize{Simplified composition overview of the granule \textit{Geolocation} with Schema.org mapping}}
\end{figure*}

After collecting the concepts and corresponding lexical items from the sources, we started  structuring the first version of the ontology, selecting classes and subclasses (step (1), figure 4). The ontology is currently available at this link: http://flux.univ-lr.fr/public/DataTourism.owl.
The mapping with Schema.org (step (2), figure 4) complements the ontology to make it more complete, up to date and coherent.
Table 2 shows the mappings between the main concepts coming from TIFSem to those in Schema.org.

\begin{table}[!h]
\centering
\caption{TIFSem main concepts/Schemas.org}
\begin{tabular}{|p{3.7cm}|p{4.3cm}|}
\hline
\textbf{CONCEPT} & \textbf{Schemas.org}\\
\hline
\textit{MULTIMEDIA} & schema.org:MediaObject\\
\hline
\textit{CLASSIFICATIONS} & schema.org:Rating\\
\hline
\textit{CONTACTS} & schema.org:ContactPoint\\
\hline
\textit{LEGAL INFORMATION} & schema.org:Organization\\
\hline
\textit{LANGUAGES} & schema.org:Language\\
\hline
\textit{GEOLOCATION} & schema.org:Place\\
\hline
\textit{RESERVATION MODES} & schema.org:Reservation, schema.org:LodgingReservation\\
\hline
\textit{PRICES} & schema.org:Offer, schema.org:PriceSpecification	\\
\hline
\end{tabular}
\end{table}

Figures 5 and figure 6 show a simplified composition overview of the two granules \textit{Geolocation} and \textit{ReservationModes}, with Schema.org mapping.

\begin{figure*}
\centering
\includegraphics[scale=0.38]{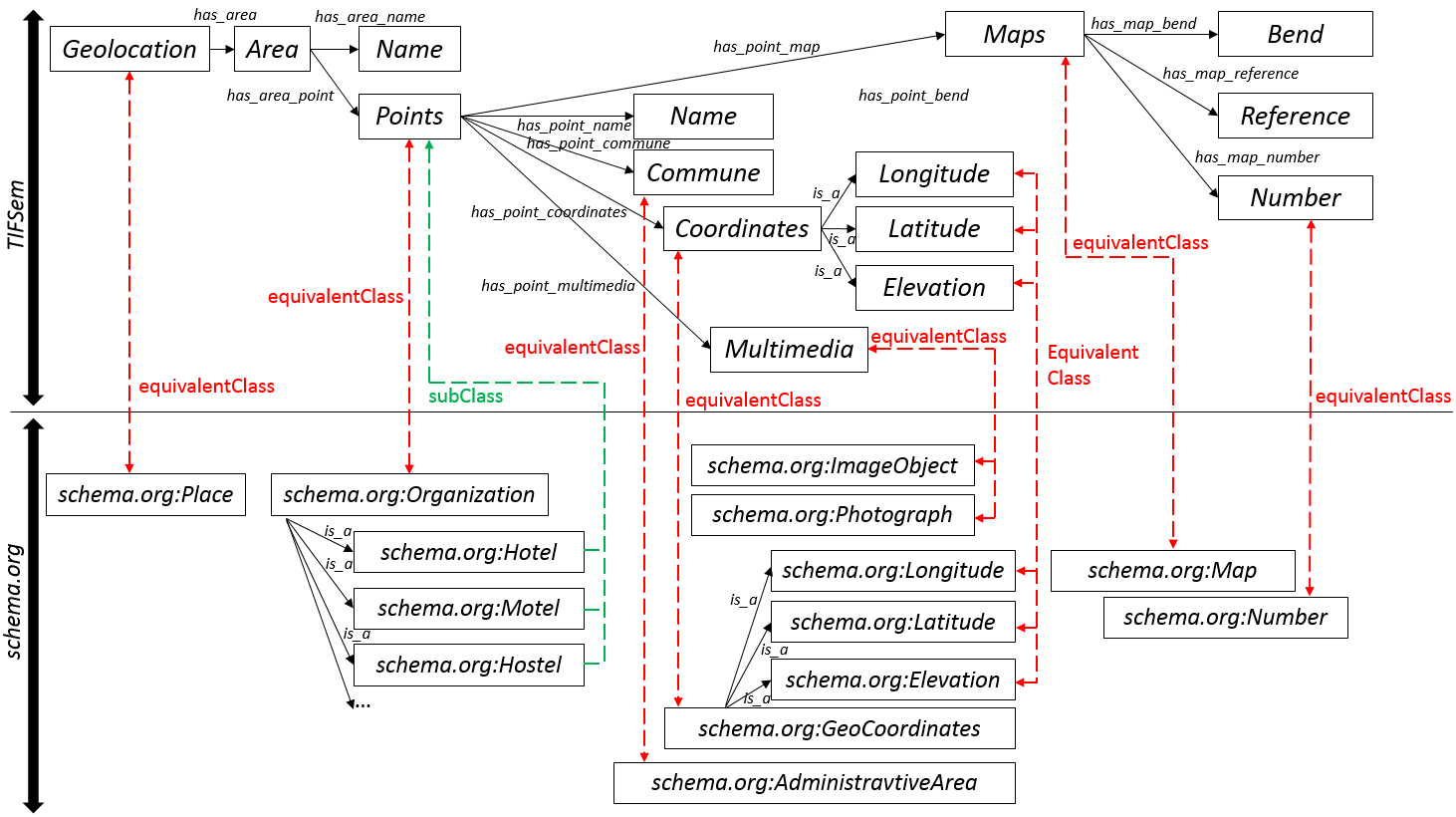}
\caption{\scriptsize{Simplified composition overview of the granule \textit{Geolocation} with Schema.org mapping}}
\end{figure*}

\begin{figure*}
\centering
\includegraphics[scale=0.38]{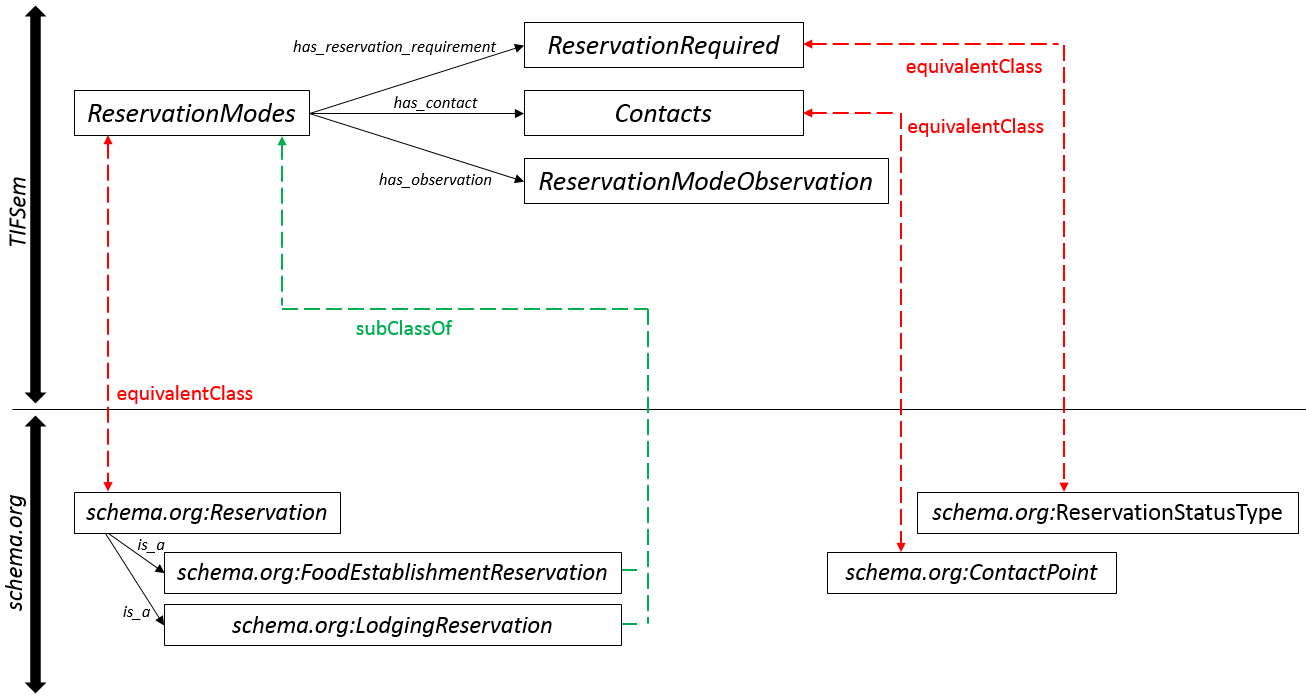}
\caption{\scriptsize{Simplified composition overview of the granule \textit{ReservationModes} with Schema.org mapping}}
\end{figure*}

The meaning of concepts is mainly expressed through the definitions in natural language.
So we used OWL language which is equipped with a rich expressive power, also to enrich Schema.org which is specified in RDF.
The use of an upper ontology, especially when there is extensive use of multiple inheritance, has the advantage of providing a manual navigation of the ontology and therefore simplify the recovery of concepts that can serve as a description.
The last step in the development process of TIFSem is the integration of instances into the ontology coming from CDT17 and CDT10 (step (3) (4), figure 4).\\
To illustrate our approach, we consider a first example that ranks different accommodation offers based on their proximity to restaurants, bars and events in La Rochelle.
A couple wants to spend a week on vacation may choose between hotels not far from downtown.
The couple likes dining out, going to bars and enjoying events.
Therefore, an additional characteristic that quantifies the aptness of each hotel for those that like to go out, would be of great help.
However, such qualitative information is rarely available on tourism on-line Web sites.
One possibility would be to compute such semantic annotations based on the geolocalisation of the hotels and those objects related to restaurants, bars and events.
Figure 7 shows a simplified SPARQL query reflecting the above example.
Distance function filters any potential results whose corresponding return value was not less than a constant value.\\
As second example, the tourist office of La Rochelle wants to have a better visibility of its territory by extracting information about rural events occurring on their territory.
They need to know the events audience and information about the public attending these events.
Figure 8 shows a simplified SPARQL query reflecting the second example.

\begin{figure}[!h]
\centering
\includegraphics[scale=0.58]{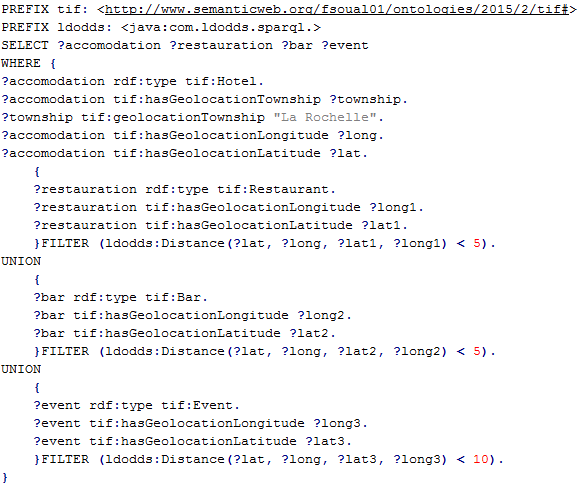}
\caption{\scriptsize{Simplified SPARQL query, example 1}}
\end{figure}

\begin{figure}[!h]
\centering
\includegraphics[scale=0.58]{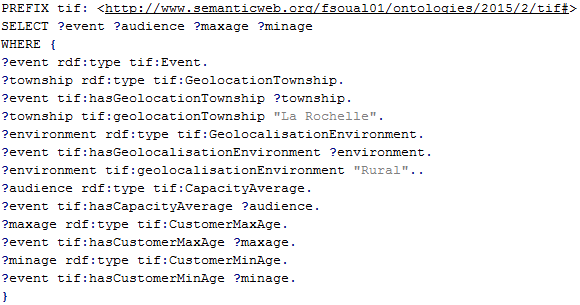}
\caption{\scriptsize{Simplified SPARQL query, example 2}}
\end{figure}

\section{Conclusions}
In this paper, we presented some early stage work in the Tourinflux project on identifying tourism domain ontology.
Semantic technologies provide methods and concepts facilitating effective integration of tourism information originating from various sources, representing basic notions and conceptual relations in tourism domain.
The objective is to improve the display of search results, making it easier for tourism industrial to find data in the right Web pages.
In particular having annotations on websites that can be understood by search engines busts the on-line visibility and increases the chances that Web sites are in the search engines results to a relevant query.
As part of our current and future work we are developing a Schema.org mapping to facilitate Web searching of tourism data.
We are also in the process of extending the TIFSem ontology by collecting contents about more touristic service providers.
We also plan to expand the TIFSem model with other types of data (about opinions, temporal data), to provide semantic and contextual answers to queries.


%
\bibliographystyle{abbrv}
\bibliography{sigproc}  
%
%
\end{document}